# NUCLEUS: A PILOT PROJECT

A secure collaboration space and data repository for Los Alamos National Laboratory Research.

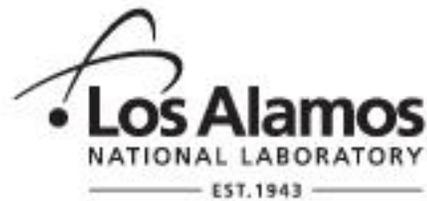

*Project Team*

Martin Klein**,** Joshua Finnell, Brian Cain, Herbert Van de Sompel,

Lyudmila Balakireva, Harihar Shankar, Jason Keith, and James Powell

*Sponsor*

Dee Magnoni

# Environmental Scan of Research Data Services: Los Alamos National Laboratory and Peer Institutions

______________________________________________________________

Early in 2016, an environmental scan was conducted by the Research Library Data Working Group for three purposes:

1.) Perform a survey of the data management landscape at Los Alamos National Laboratory in order to identify local gaps in data management services.
2.) Conduct an environmental scan of external institutions to benchmark budgets, infrastructure, and personnel dedicated to data management.
3.) Draft a research data infrastructure model that aligns with the current workflow and classification restrictions at Los Alamos National Laboratory.

The first phase of data collection consisted of 24 in-depth interviews with researchers from across the Lab and were completed during the summer of 2016.

1. **Data Interviews** - The individuals who were interviewed spanned a diverse set of divisions, positions, and career stages but should not be considered comprehensive. It should also be noted that previous data collection in the area of data management has been undertaken at the Lab in the last five years. Surveys were conducted or attempted by the Research Library in 2011 and by a previous iteration of the Data Working Group in 2015. Also in 2015, Reid Priedhorsky conducted data management interviews with members from High Performance Computing. These surveys and interviews provided both historical context and guideposts in conducting this current version of data interviews.

   Moreover, this survey confirmed that data management needs remain consistent at the Lab across the last five years in terms of data management planning, data storage, and data collaboration and dissemination.

   - **Data Management Planning** – The awareness and completion of data management requirements and mandates was limited in our survey pool, even with the Department of Energy's Office of Science mandate requiring data management plans (DMPs) for federally-funded research on October 1, 2015. The few interviewees who created a data management plan used the library-sponsored DMPTool, but better support and services were identified in this area.

   - **Data Storage** – Echoed in the data surveys conducted in 2011, the lack of a centralized data storage solution at the Lab, that meet the needs of the research community, was a common theme. For myriad reasons, from mere efficiency to cost, the most commonly employed approach to data storage and preservation is a personal computer or local network drive.

- Data Collaboration – Connected to the issue of centralized storage is a desire among many respondents for collaborative tools (i.e. Google Drive, Dropbox) to work with lab partners on research projects. Currently, LANLTransfer or email attachment is the preferred method of sharing data with researchers both internally and externally

- Dissemination. Additionally, the majority of researchers requested assistance with submitting their data through RASSTI, as the current system requires copying data to a physical CD and delivering it to SAFE-1 for review. Though time constraints in securing funding and conducting research was often cited as a challenge to data management by researchers at the lab, the lack of a centralized data repository and effective collaborative infrastructure was also identified an obstacle in fulfilling funder and publisher mandates as well as conducting scientific research efficiently and effectively.

The second phase of data collection consisted of contacting data librarians and managers at 12 institutions, universities and national laboratories, during the spring of 2016.

2. Institutional Benchmarking – The majority of institutions surveyed have an average operating budget of $500,000 annually dedicated to data management services. These budgets support on average 3 full-time employees and a data platform such as Dataverse or DSpace. Collaboration was a common theme across institutions, with data services being executed collaboratively between 3 or more departments (see Appendix 1).

## Goals and objectives

In response to these findings, the proposal is to provide institutional infrastructure that facilitates management of research projects, research collaboration, and management, preservation, and discovery of data. Deploying such infrastructure will amplify the effectiveness, efficiency, and impact of research, as well as assist researchers in regards to compliance with both data management mandates and LANL security policy. This will facilitate discoverability of LANL research both within the lab and external to LANL. Paramount to this proposal is helping researchers throughout the entire lifecycle of their research by:

1) Facilitating sharing to help ameliorate the group, institutional, and disciplinary barriers that inhibit true cross-collaboration on research projects

2) Integrating tools that are commonly used for research purposes into the proposed platform to ensure utility
3) Providing an integrated, seamless, and uncomplicated platform for effortless compliance with all LANL data management, review and release, and security policies
4) Improving documentation and preservation of research data to prevent institutional memory loss
5) Providing a straightforward method for LANL researchers to fulfill funder and publisher mandates regarding data management

The Research Library proposes a pilot project to implement a prototype of such institutional infrastructure by combining existing components. The prototype infrastructure will illustrate its usability across the entire data life cycle, as illustrated in the below diagram.

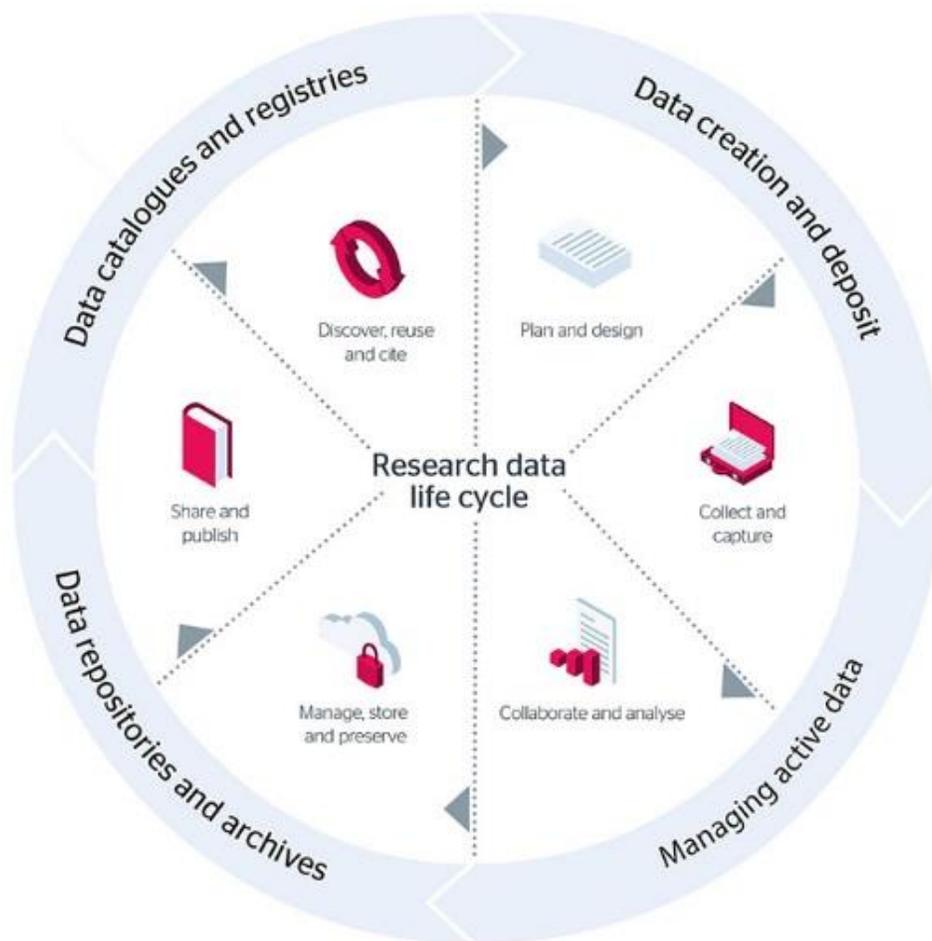

Figure 1: A research data life cycle diagram[1]

---

[1] JISC. https://www.jisc.ac.uk/guides/meeting-the-requirements-of-the-EPSRC-research-data-policy

# 3. Nucleus: A Collaborative Data Platform and Repository

This proposal lays out a one-year plan of work to be conducted at the Research Library, leading to the operational deployment of a pilot project known as Nucleus. The main components of this platform will consist of:

1. A collaborative layer for active projects
2. Sync and share storage
3. Integration with the Data Management Plan Tool
4. Integration with review and release protocols (RASSTI Data)

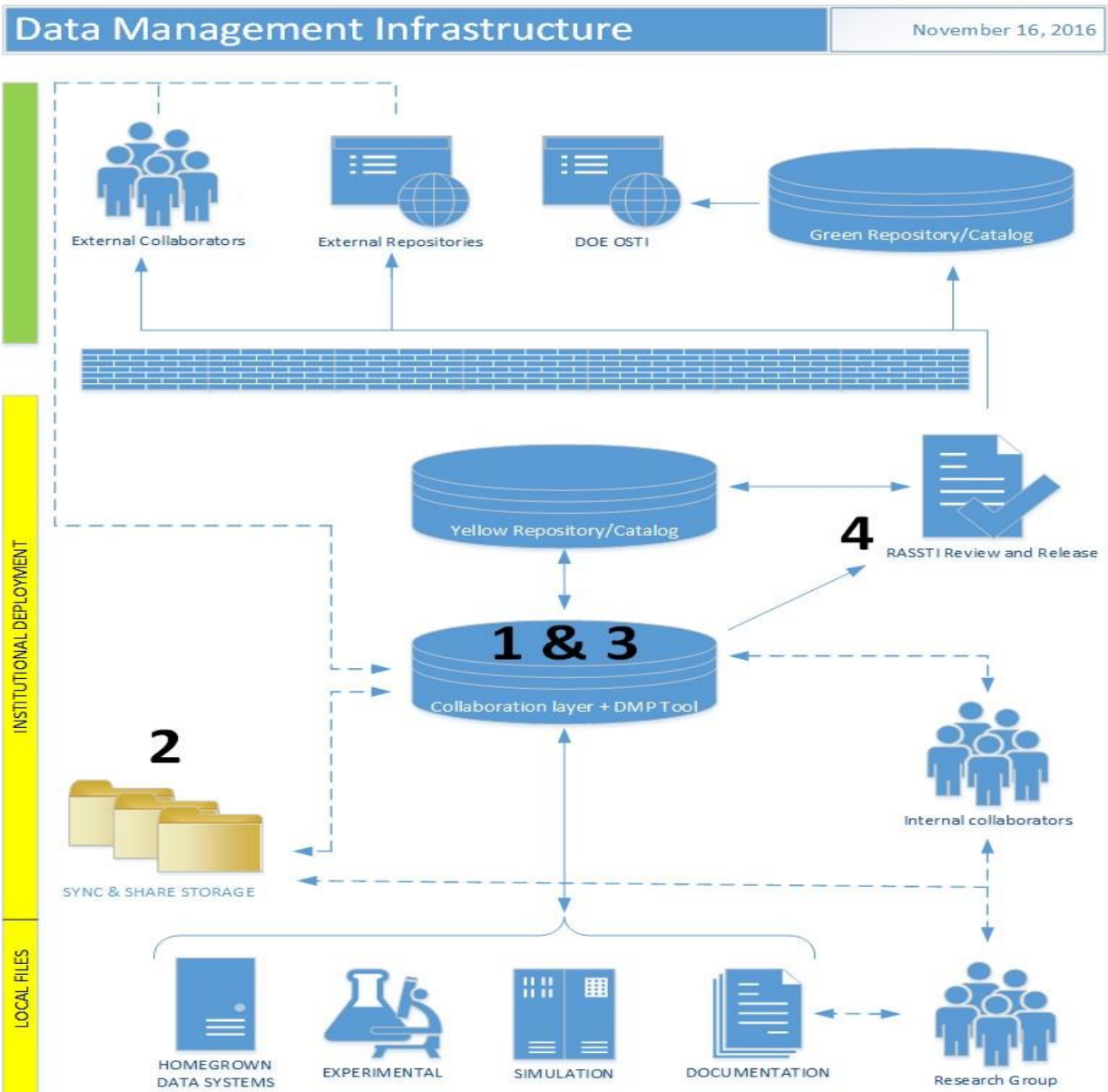

# 1. Collaboration layer

Software solution: local Open Science Framework

Developed by the Center for Open Science, the Open Science Framework is an open source software project that simplifies collaboration in scientific research. This solution is meant to facilitate the bringing together of disparate things in one central location. Ideally, this space can integrate many different sources of data under the umbrella of a project including resources from a local ownCloud instance, a local Gitlab/Github instance, a homegrown data management system (eg: Echo, MDS+, Granta, etc.), statistical or analysis software output (R, Python, MATLAB, etc.), lab notebook/workflow/vizualization software (Jupyter, Kepler, VisTrails, etc.), local repository solutions, direct uploads from users, and other custom niche applications research groups use which are all aggregated and accessible through a web interface. Furthermore, these project spaces have permissions associated with them, allowing researchers to add individuals to projects for read, read-write, or full administrative privileges. Other features include file version control, wikis, analytics, file previews, "cloud" storage, search and discovery, and archiving capabilities. This collaboration layer will provide a high degree of functionality relevant to a research group, but also be elegant in its simplicity to bring together variegated data sources and collaborators. As a result, the likelihood of collaboration, data sharing, and institutional memory retention will be greatly amplified. After a preliminary analysis of suitable tools available, the Open Science Framework (http://osf.io) appears to be the most robust solution for this collaboration layer.

# 2. Sync and Share Storage

Possible software solution: ownCloud; Minio

Open source software solutions can be deployed on LANL machines with similar functionality as cloud storage solutions like Dropbox. This is not true cloud storage as it does not run on servers external to the lab like Dropbox or Box. Permissions allow for creating public and private storage folders that can be shared among researchers using a desktop client or web interface. Encryption is also available for an added layer of security.

There are numerous benefits to deploying an institutional sync and share cloud service:

1. Researchers will easily be able share working folders with themselves and others in a device agnostic manner (Desktop, Mobile, etc.)
2. Assurance that important research materials are backed up in multiple locations
3. Collaborative online document editing
4. Less likely that data will be lost when researchers leave the lab
5. Reduced reliance on random physical media (external hard drives, flash drives, DVDs) which are often unlabeled, misplaced, and become unreliable/depreciated over time
6. HTTP URI namespace enabling access to versions of resources

## 3. Data Management Plan Tool (DMPTool)

Since 2014, the Research Library has provided researchers access to the DMPTool, an online tool that includes data management plan templates for many of the large funding agencies that require such plans, including the Department of Energy. Building upon the work of the California Digital Library, the research team will integrate the DMPTool into the local Open Science Framework platform to allow seamless integration from data management planning to workflow and deposit.

There are numerous benefits to integrating the DMPTool into localized Open Science Framework:

1. Creates readily available reference source in a centralized platform for researchers seeking information about research sponsor requirements.
2. Centralizes data management planning into the existing workflow of researchers.
3. Creates a "living" document that can be modified as data changes during course of active research.
4. Facilitates the deposit of data management plan along with data for future researchers to understand and reuse the data.

## 4. RASSTI Data – review and release

Software solution: Homegrown system maintained by the library

A RASSTI Data system will be deployed in order to accommodate research data. Original RASSTI was designed to review traditional publications or presentations. Research data is often bundles of multi-modal files, consisting of everything from text files to complex scripts and software. RASSTI's inability to handle file formats beyond a PDF is problematic for researchers wanting to release their data to publications, external repositories, or non-LANL collaborators. The deployment of technologies outlined above and a RASSTI Data system will allow for tighter integration and nearly effortless methods for researchers to have research data reviewed and released. With the removal of this barrier and integration into a larger workflow platform the Research Library wishes to pilot, RASSTI Data will:

1. Simplify the process of submitting data for review and release
2. Improve security compliance and reduce hurdles when engaged with external collaborators
3. Increase amount of open LANL research data available from DOE's SciTech discovery portal
4. Improve discoverability of LANL research data through enhanced metadata application to datasets

# Timeline

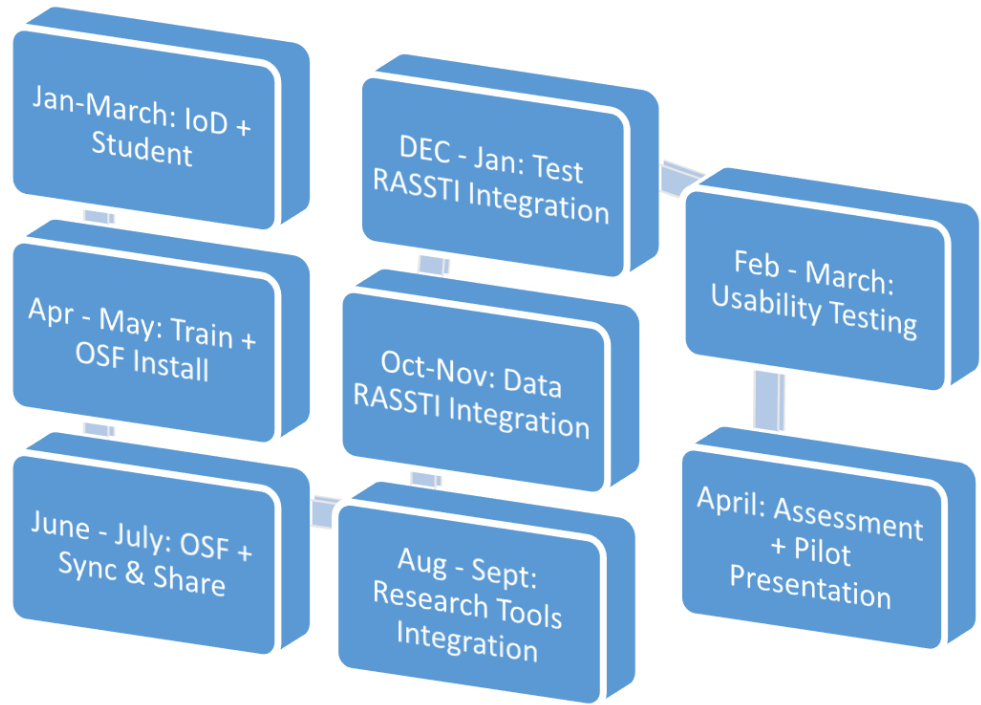

This project consists of approximately11 months of prototyping and technical review and 1 month of testing.

## 2017

**January - March**: Identify and hire qualified student + IoD Server

**April - May**: Train student + OSF installation

**June - July**: Integration and testing of OSF and ownCloud/Minio/Sync & Share onto IoD server

**August - September:** Explore integration of research tools: DMP Tool, GitLab, etc.

**October - November:** Explore RASSTI Data integration

**December - January:** Test RASSTI Integration + Holiday Break

## 2018

**February – March:** Usability testing with selected lab group (EES-16, Center for Nonlinear Studies)

**April:** Presentation + Final Report + Assessment of Next Steps + Student Hire Expiration

# Project Evaluation

## 1. Viability of Localized Infrastructure

The installation of Open Science Framework and integration of ownCloud/Minio onto a localized server are foundational to the success of this project. Successful installation of this software in a localized environment will be assessed by its functionality compared to its cloud-based equivalent and sustainability in handling local patches and security updates. A simulated, text project in both environments will be performed and documented to provide a viability report. Moreover, the integration of each additional research tool (GitLab, for example) will be tested and documented for functionality and sustainability.

## 2. Usability with Stakeholders

Identified during the data interviews, a small group of researchers in EES-16 will serve as our usability testers, assessing the usability of the platform within their existing worfklows. Members of this group will then participate in a focus group to provide feedback on improvements and functionality. The same opportunity will be extended to members in the Theoretical Division of the laboratory. These two cohorts represent both traditional and non-traditional data users in the lab, and will provide an initial assessment of the platform's ability to be deployed across a diverse group of divisions and programs.

# Budget

Because this pilot project is limiting its scope of inquiry to open-source platforms, there is no cost for software licenses. Initial deployment of this pilot project will be installed on the Lab's Infrastructure on Demand server, requiring no additional funding. Members of the three collaborative teams in the Library (Customer Engagement Team, Institutional Scientific Content Team, and Digital Library Research and Prototyping Team) will provide their labor out of regular operating budgets. The post-baccalaureate student's salary will be provided through the Digital Library Research and Prototyping Team.

The extensibility and sustainability of this project beyond the pilot project will require an investment of both people and budgeting.  The Institutional Benchmarking Report of similar institutions provided in Appendix 1, provides an estimate of both personnel (2-3 FTE) and budget ($300,000 approximately) necessary to deploy this platform across the Lab.

# Project Team

Martin Klein - Digital Library Research and Prototyping Team, SRO-RL

Joshua Finnell – Customer Engagement Team, Data Working Group, SRO-RL

Brian Cain – Institutional Scientific Content Team, Data Working Group, SRO-RL

Herbert Van de Sompel – Digital Library Research and Prototyping Team, SRO-RL

Lyudmila Balakireva - Digital Library Research and Prototyping Team, SRO-RL

Harihar Shankar - - Digital Library Research and Prototyping Team, SRO-RL

James Powell - - Digital Library Research and Prototyping Team, SRO-RL

Jason Keith – Post-Master's Student, Digital Library Research and Prototyping Team, SRO-RL

# Institutional Data Benchmarking Summary Table

| Institution | Budget | Staffing | Platform | Discovery Tools | Cooperation | Permanent Storage |
|---|---|---|---|---|---|---|
| Imperial College London | Asked for $300K; Received $15k | 5 FTE | DSpace/Box/OneDrive | Symplectic | Research Office + IT + Library | N/A refer users to Zenodo |
| Johns Hopkins University | $10,000,000 Grant from NSF; Annual not disclosed | 6 FTE + 1 PT + 3-5 students | Dataverse | DOIs + APIs + Visualizations | Digital Research and Curation Center + Entrepreneurial Program + Library + IT + Research Administration Office + Biostats Center | Data Conservancy |
| Lawrence Livermore National Laboratory | N/A | 1 PT | Drupal | OSTI Harvest | Program Development Office + Library + IT + Information Mgmt Group | None/External Storage Recommended |
| McGill University | $500K Annual | 4FTE + 5 PT | None; Exploring Dataverse + Hydra | None | Library + IT + Office of Research and International Relations Office. | Exploring Dataverse + Archivematica |
| National Renewable Energy Laboratory | $300K Annual | 3FTE + 1 PT | Homegrown; AWS | None | Publications Office + Data Analysis + Visualization Group + HPC | None |
| Oak Ridge National Laboratory | N/A | 3FTE | Homegrown; CADES System | None | Computing + Library | None |
| Purdue University | $1,200,000 Institutional Grant; $330K Annual | 5FTE + 2-3 students | Homegrown; PURR/HubZero | DOIs + APIs | Library + Office of Research + IT + Executive Committee + Faculty Committee + Archives | 10-year retention policy; LOCKSS |
| University of California, San Diego | $2,000,000 Annual | 10 FTE | Hydra/Fedora; Chronopolis; Qumulo | Data Catalog | Library + IT + Scripps Institute for Oceanography | None |
| University of Maryland | $200K Annual | 2 FTE + 1 student | DSpace | DOIs | Library + IT + Office of Sponsored Research | None |
| University of Michigan | $750K Annual | 3 FTE + 3 students | DSpace + Hydra/Fedora | DOIs + APIs | Library + IT + Office of Sponsored Research | None; 10-year retention policy |
| University of New Mexico | $200K Annual | 2 FTE + 3 students | ownCloud + GitHub + DSpace/ exploring Bepress | DOIs + APIs | Library + IT + Office of Innovative Scholarly Initiatives | None; 10-year retention policy; exploring Data Preservation Network |
| University of Waterloo | N/A | 3 PT | Dataverse | Primo Indexing | Library + IT + Office of Research | Archivematica |
| | Annual Range: ($200K -$2,000,000) | Staffing Range: (2-6 FTE) | Common: (Hydra/Fedora + Dataverse + DSpace) | Common: (DOIs + APIs) | Common: (partnership between 3+ departments) | Common: (10-year retention policy) |

Appendix 1